\documentclass{actasen}

\begin{document}

\setcounter{page}{00}

\Volume{0000}{00}


\runheading{LIN Meng-xiang}

\title{Oscillation: The Key For Understanding Strange Radio Behaviors Of AXP/SGRs}

%
%
%
%

\footnotetext{
\hspace*{5mm}$^{\dag}$ linmx97@gmail.com\\
}

\enauthor{LIN Meng-xiang$^{\dag}$ }{Department of Astronomy, Peking University, Beijing 100871, China.}

\abstract{
We suggest stellar oscillations are responsible for the strange radio behaviors of Anomalous X-ray pulsars and soft Gamma-ray repeaters (AXP/SGRs), within the framework of both solid quark star model and magnetar model.
In solid quark star model, the extra voltage provided by oscillations activates the star from under death line to above death line. In magnetar model, oscillations enlarge the radio beam so that increase the possibility to detect it. Later radio emission decays and vanishes as oscillations damp.
}


\maketitle

\section{Introduction}

In pulsar family, anomalous X-ray pulsars and soft gamma-ray repeaters (AXP/SGRs) are quite special. They are found from X-ray and most of them are radio quiet. They are also thought to be magnetars which was first presented by {Tompson \& Duncan}\rf{1,2}. In magnetar model, AXP/SGRs could have superstrong magnetic field up to about $10^{14}{\rm G}$). But for radio death line criterion, such superstrong magnetic field should make them radio loud. So, the superstrong magnetic field was thought to be capable to suppress the radio emission. However, up to now, we indeed have detected four radio AXP/SGRs \rf{3}. In quiescent state, they are radio quiet; but after their X-ray outbursts, they became radio loud, and the radio emission decayed in a timescale of months or years. Why are AXP/SGRs' radio behaviors so strange? We still don't have a good understanding.

From observation, AXP/SGRs' X-ray outbursts are sometimes associated with glitches (sudden "spin-up" of the star) \rf{4}, and we could expect that stellar oscillations would be excited after glitches. Therefore, we would like to explore the oscillations' effect on the radio behaviors. In the other hand, the physical nature of AXP/SGRs is still an open question. The prevalent model is magnetar model, thinking they are neutron star with superstrong magnetic field. Alternatively, a solid quark star model was suggested by Xu \rf{5,6}. Here we discuss radio emission behaviors of AXP/SGRs within both models.

Lin et al. \rf{7} systematically studied the stellar oscillations' effect on magnetospheric activity within the framework of inner gap model \rf{8}, and applied this physical model to explain the radio behaviors of AXP/SGRs and glitch-induced radio profile change of PSR J1119-6127. In this letter, based on Lin et al. \rf{7}, we mainly focus on the work about radio behaviors of AXP/SGRs and give a brief summary to it.

\section{OSCILLATIONS' EFFECT}
When a magnetized object rotates, it will produce electric field around it due to unipolar induction. Pulsars are exact such objects. Since their strong magnetic field, they could produce strong electric field in the surface, and the electric field may have the ability to accelerate charged particle, thus may produce radiations. Now we consider toroidal oscillation. Similar to rotation, toroidal oscillation also could produce extra electric field due to unipolar induction, thus influence the particle acceleration and radiation.

Now we consider toroidal oscillations' effect on inner gap model \rf{8}. In inner gap model, due to large bound energy of pulsar surface, there will be a vacuum gap above the surface, and charged particles are accelerated within this vacuum gap. An important physical quantity here is the voltage that could be used to accelerate the charged particles. In order to emit radio emission, a pulsar should provide large enough voltage. The most significant effect of stellar oscillations is to produce extra voltage.

\section{EXPLANATION OF RADIO BEHAVIORS OF AXP/SGRS}
\subsection{In Solid Quark Star Model}
In solid quark star model, the superstrong magnetic field is not necessary, thus AXP/SGRs could have magnetic field comparable to normal radio pulsars ($\sim 10^{12}{\rm G}$). Therefore, they could naturally be radio quiet in their quiescent states. After X-ray outbursts, stellar oscillations are excited, thus the extra voltage provided by oscillations could activate the star to be radio loud. With the oscillations damp away later, radio emission decays and finally vanishes.

In figure 1, using the parameters of a detected radio AXP/SGR, 1E 1547.0-5408, assuming a $10^{12}{\rm G}$ dipole magnetic field, we calculate two qualities as functions of oscillations' amplitude, says the voltage along the vacuum gap $\Delta V_{\rm gap}$ and the maximum voltage that the star could provide $\Delta V_{\rm max}$. $K$ presents the dimensionless amplitude of oscillations, which approximately equals to the ratio of oscillation velocity and rotational velocity in the star surface.
A necessary condition for radio emission is that the star should be above the radio death line, which means the voltage along the gap should be larger than the maximum voltage the star could provide. When the oscillation amplitude $K$ is small, the star is under the death line and thus radio quiet; after $K$ exceeds a critical value, the star would be activated above the death line, and thus could be ratio loud.
Table 1 in Lin et al.\rf{7} shows such critical values for all detected radio AXP/SGRs.

\begin{figure}
    \centering
    \includegraphics [angle=0,width=9cm]{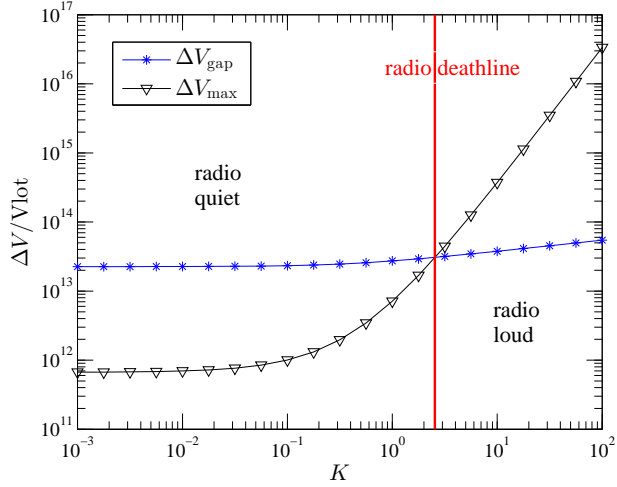}
    \caption{Voltage along gap ($\Delta V_{\rm gap}$) and the maximum voltage ($\Delta V_{\rm max}$) as functions of dimensionless amplitude of oscillations ($K$). When $\Delta V_{\rm gap}$ is larger than $\Delta V_{\rm max}$, the star is under death line{\bf, thus radio quiet}.}
\end{figure}

\subsection{In Magnetar Model}
In magnetar model, AXP/SGRs have superstrong magnetic field, so they are always above the death line. However, rather than X-ray, radio emission have strong beaming effect. Therefore, we may miss them radio emission if the radio beam don't sweep over the line of sight. In the other hand, besides provide extra voltage, oscillation could also enlarge the polar cap region, thus enlarge the radio beam and increase the possibility to detect them. Same as that in solid quark star model, oscillation damping would make radio emission decay and vanish.

\acknowledgements{The author would like to express hearty thanks to organizer of QCS2014 and the helpful discussion of PKU Pulsar Group.}

\end{document}